\documentclass[conference]{IEEEtran}
\usepackage{graphicx}
\usepackage{fancyhdr}
\usepackage{fancyhdr}

\begin{document}
\title{An Effective Private Data storage and Retrieval System using Secret sharing scheme based on Secure Multi-party Computation}
\author{\IEEEauthorblockN{Divya G Nair}
\IEEEauthorblockA{Department of Computer Science\\
Cochin University \\
Kochi, India\\
divyagnr@gmail.com}
\and
\IEEEauthorblockN{Binu.V.P}
\IEEEauthorblockA{Department of Computer Application\\
Cochin University\\
Kochi, India\\
binuvp@gmail.com}
\and
\IEEEauthorblockN{G.Santhosh Kumar}
\IEEEauthorblockA{Department of Computer Science \\
Cochin University\\
Kochi, India\\
sancochin@gmail.com}}
\fancypagestyle{plain}{
%\fancyhf{}	% clear all header and footer fields
\fancyfoot[L]{\vspace{-12mm}\footnotesize{978-1-4799-5461-2/14/\$31.00~\copyright 2014 IEEE}}
\fancyfoot[C]{2014 International Conference on Data Science \& Engineering (ICDSE)}
\fancyfoot[R]{}
\renewcommand{\headrulewidth}{0pt}
\renewcommand{\footrulewidth}{0pt}
}

\pagestyle{fancy}{
%\fancyhf{}
\fancyfoot[C]{2014 International Conference on Data Science \& Engineering (ICDSE)}
\fancyfoot[R]{}}
\renewcommand{\headrulewidth}{0pt}
\renewcommand{\footrulewidth}{0pt}
% make the title area
\IEEEoverridecommandlockouts
\IEEEpubid{\makebox[\columnwidth]{978-1-4799-5461-2/14/\$31.00~\copyright 2014 IEEE \hfill} \hspace{\columnsep}\makebox[\columnwidth]{}}

\maketitle
\thispagestyle{plain}

\maketitle
\begin{abstract}
%\boldmath
Privacy of the outsourced data is one of the major challenge.Insecurity of the network environment and untrustworthiness of the service providers are obstacles of making the database as a service.Collection and storage of  personally identifiable information is a major privacy concern.On-line public databases and resources pose a significant risk to user privacy, since a malicious database owner may monitor user queries and infer useful information about the customer.The challenge in data privacy is to share data with third-party and at the same time securing the valuable information from unauthorized access and use by third party.A Private Information Retrieval(PIR) scheme allows a user to query database while hiding the identity of the data retrieved.The naive solution for confidentiality is to encrypt data before outsourcing.Query execution,key management and statistical inference are major challenges in this case.The proposed system suggests a mechanism for secure storage and  retrieval of private data using the secret sharing technique.The idea is to develop a mechanism to store private information with a highly available storage provider which could be accessed from anywhere using queries while hiding the actual data values from the storage provider.The private information retrieval system is implemented using Secure Multi-party Computation(SMC) technique which is based on secret sharing.  Multi-party Computation enable parties to compute some joint function over their private inputs.The query results are obtained by performing a secure computation on the shares owned by the different servers.

\end{abstract}
\section{Introduction}
% no \IEEEPARstart

Secure storage of confidential data and their private retrieval are major research challenges, when the data are outsourced to a third party untrusted service provider.Private Information Retrieval(PIR) allows clients to retrieve data from a database server in a privacy-preserving manner.PIR schemes make use of cryptographic protocols to safeguard the privacy of database users. This allow clients to retrieve records from public databases, while the identity of the retrieved records is completely hidden from database owners. The major goal is that the database server should be able to respond to client queries without learning any information about the records retrieved.

A trivial solution is to encrypt the database \cite{zhu2007executing} using cryptographic techniques.But for the query processing, the entire database must be downloaded  and  queries must be issued locally.The query execution over encrypted data is a major research challenge.Most of the solutions are inefficient due to the large query processing time and complexities involved in key management. The use of  encrypted database  is clearly information-theoretically secure and the server cannot learn which record the client seeks, but the key management, time consuming encryption decryption process, overhead in large encrypted database downloading and the difficulties involved in query processing make the scheme impractical. 
 
Fragmentation is another solution for providing confidentiality of the outsourced data \cite{wiese2010horizontal}.The data owner partitions the tables horizontally or vertically and distribute them to different servers.Encryption is unavoidable in this case also because fragmentation cannot preserve the confidentiality of a single attribute.Collusion between servers is also a security issue.Agarwal et al \cite{agrawal2009database} use secret sharing technique to provide confidentiality.Their solution supports different type of queries to run efficiently.But untrusted servers have prior knowledge about data distribution or frequency.
  
The proposed system suggests a secret sharing method for confidentiality in the outsourced data.A relation is split into random shares and these shares are send to the different servers.This provides both reliability and security.The threshold secret sharing scheme helps the data to be retrieved from $k$ number of servers out of $n$ servers where the shares are stored.The random shares also provides information theoretical security at the cost of additional storage space. Query processing and searching is an issue here.An efficient mechanism for searching and query processing is also suggested in this paper.It needs interaction between client and different servers.The servers will send the shares which are the results of the query.The shares are then combined to form the original data.Since the computations are performed on shares,
it provides a Secure Multi party Computation(SMC) environment.

 There are several situations in which  mutually distrustful parties need to perform a joint computation without revealing their inputs to each other. This happens, for example, during auctions, voting, negotiations and business analytics. The problem is how to perform such a computation without revealing the inputs.SMC\cite{goldwasser1997multi} is the solution to such problems.It permits a group of parties to jointly compute a function of their private inputs while preserving  privacy and correctness of input.Every participant will get the result of computation without exposing their input.SMC protocol was first introduced by Yao in 1982 by exploring the famous Millionaire's problem\cite{yao1982protocols}. The protocol is secure, if no participant can learn more from the description of the public function and the result of the computation. 
	
SMC is accomplished here by using Shamir's secret sharing scheme. In secret sharing, the secret is not single handed, but multi-handed so that even if any of the parties involved in the computation are malicious, the secret can be reconstructed. A  verifiable secret sharing scheme is one in which  parties can verify the validity of the shares for consistency. To handle malicious parties involved in any computation, the secret sharing scheme needs to be verifiable. 
	
Development of secret sharing scheme started as a solution to the problem of safeguarding cryptographic keys by distributing the key among $n$ participants and $t$ or more of the participants can recover it by pooling their shares. Thus the authorized set is any subset of participants containing more than $t$ members.This scheme is denoted as $(t,n)$ \textit{threshold scheme}\cite{desmedt1990threshold}. The notion of a threshold secret sharing scheme is independently proposed by Shamir\cite{shamir1979share} and Blakley\cite{blakley1899safeguarding} in 1979. Since then much work has been put into the investigation of such schemes. Linear constructions were most efficient and widely used. A threshold secret sharing scheme is called \textit{ideal}, if the share size is same as the secret size and is \textit{perfect}, if less than $t$ shares give no information about the secret.Blakley's scheme is not perfect while Shamir's scheme is perfect. Both  Blakley's and the Shamir's constructions realize $t$-out-of-$n$ threshold secret sharing scheme. However,their constructions are fundamentally different.
	
Shamir's scheme is based on polynomial interpolation over a finite field. It uses the fact that we can construct a polynomial of degree $t-1$ only if  $t$ data points are given. A polynomial $f(x)=\sum_{i=0}^{t-1}a_ix^i$, with $a_0$ is set to the secret value and the coefficients $a_1$ to $a_{t-1}$ are assigned random values in the field, is used for secret sharing.The polynomial $f(x)$ is evaluated at $n$ different points and each value is given as share to a participant. That is , the value $f(i)$ is given to the user $i$ as secret share.Here $t$ is considered as threshold. When any $t$ out of $n$ users join together they can reconstruct the polynomial using Lagrange interpolation with $t$ points and hence obtain the secret  $a_0$. Any set of $t-1$ users cannot gain any information about the secret and is a perfect scheme. This scheme is easily computable when necessary data is available and it avoids single point of failure . Also it increases reliability, security, safety and convenience \cite{binu2014epitome}.
	
The rest of the paper is organized as follows.Related works are given in section II.The proposed system and architecture are mentioned in section III.Section IV contains an example.Conclusions are drawn in section V.

\section{Related work}

The PIR (Private Information Retrieval) was introduced by Benny Chor \cite{chor1998private} and has already received a lot of attention\cite{cachin1999computationally}\cite{ishai1999improved}\cite{woodruff2005geometric}\cite{gentry2005single} . The study of PIR is motivated by growing concern about the user's privacy when querying a large commercial database .
Protocols for PIR \cite{chor1998private} and Symmetric Private Information Retrieval (SPIR)\cite{saint2005java} provide a limited type of privacy preserving search. In PIR the server and clients are involved, where the server has a database of $n$ items and the client wants to obtain the item at position $i$ without the server learning the value of $i$. In the case of SPIR, it is additionally required that the user does not learn any information about other item except the one that was requested. These protocols  improved the general multiparty computation and have sub-linear communication and polynomial computational complexity.But still these protocols remain inefficient for many practical uses and  support only simple selection, rather than general query capability.\\
In database outsourcing, one party possess large amount of data, but does not have enough storage at hand for the reliable
data storage.Many papers address the issues related with  database outsourcing \cite{boneh2004short}\cite{williams2008building}\cite{ceselli2005modeling}.The major issue is that we have to keep the data confidential from untrusted server and it must be retrieved without revealing any info. The approaches of \cite{ceselli2005modeling}\cite{saint2005java}	use encryption systems.The searching over encrypted data is time consuming,
which need the word to be encrypted before searching.Thus the running time of the search in these approaches is linear in the number of all searchable tokens and the searching become inefficient even though it provides better security. This pinpoints the issue for trade off between efficiency and strong privacy guarantees. Curtmola et al.\cite{chor1998private} use the idea of inverted indices for efficiency gain.They suggest  preprocessing of  the data by the querier and compute inverted indices on search words. An untrusted server can learn search pattern over multiple queries in this case.\\

SMC based on homomorphic public-key encryption  is also proposed in \cite{goldwasser1997multi},\cite{ishai1999improved}.In this each party distributes encryptions of its private inputs to the other parties. The computations are performed on this encrypted data. The homomorphic property of encryption can be used to achieve a specific functionality.Authorized set of users can do threshold decryption and the final result can be obtained.
 
\section{Proposed System}
The proposed system suggests a method of storing and retrieving private data in a secure and effective manner. The private data include personal information, sensitive information or unique identification etc. The data storage may be a private information storage using cloud database. 
\subsection{Secure Data storage}
\begin{figure}[h]
		      \centering
		      \includegraphics[width=20pc,height=12pc]{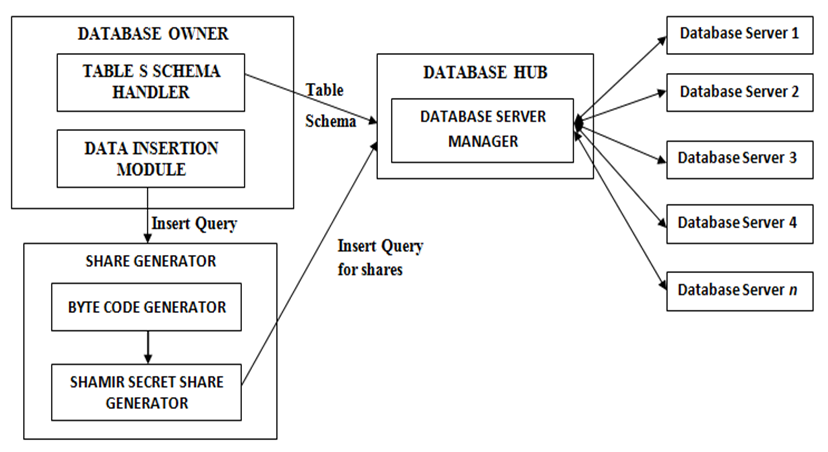}
		      \caption{Architecture for Data storage} 
		      \label{storage}
			\end{figure}

The system does not use any encryption technique.Shares corresponds to each relations are generated  using Shamir's secret sharing scheme.These shares are then stored on different servers. The architecture for data storage is shown in figure \ref{storage}

The architecture has four main modules
\begin{enumerate}
\item Database Owner
\item Share Generator
\item Database Hub
\item Database Servers
\end{enumerate}	
Database owner gives the table schema to Table schema handler which is copied into all the database servers. When a data is inserted, the data insertion module will give the data into share generator module. The share generator converts the data into a bytecode form using bytecode generator and then it is divided into shares using shamir secret share generator.The shares are then distributed and stored in different database servers.

\subsection { Algorithm for data storage}
Step1:  The database schema is copied into $n$ database servers\\
Step2: When an insert operation is performed, each attribute value is divided into shares and  stored in the database servers\\
$DS_{1}$=$R(A_{11}$,$A_{21}$,$A_{31}$,$\cdots$$ A_{m1})$\\
$DS_{2}$=$R(A_{12}$,$A_{22}$,$A_{32}$,$\cdots$$ A_{m2})$\\
$DS_{n}$=$R(A_{1n}$,$A_{2n}$,$A_{3n}$,$\cdots$$ A_{mn})$\\
where, $DS_{n}$ is the $n^{th}$ database server \\
$R(A_{1n}$,$A_{2n}$,$A_{3n}$,$\cdots$$ A_{mn})$ is the record containing attribute shares of $m$ attributes and $A_{mn}$ is the $n^{th}$ share of $m ^{th}$ attribute.\\
Step3:Along the attribute values a primary key column containing index values starting from 1 will also be created automatically.The purpose of  this is to make the retrieval process easy.
\subsection{Secure Data Retrieval}
\begin{figure}[h]
		      \centering
		      \includegraphics[width=22pc,height=12pc]{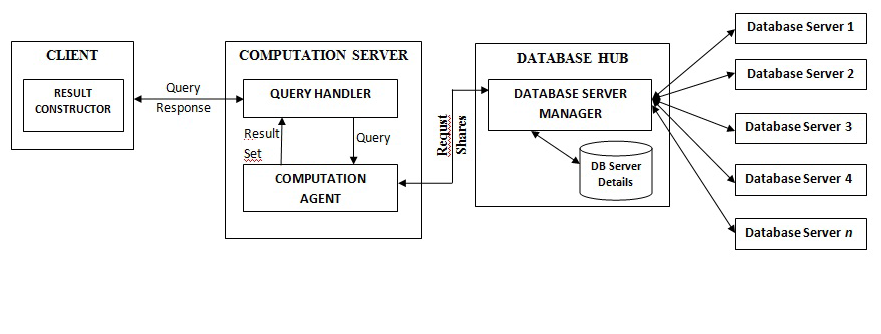}
		      \caption{Architecture for Secure Data Retrieval} 
		      \label{retrieval}
			\end{figure}

The architecture for secure data retrieval is shown in figure\ref{retrieval}\\
The main modules are \\
\begin{enumerate}
\item Client
\item Computation Server
\item Database Hub
\item Database Servers
\end{enumerate}
Step1:Client gives a query\\
Step2:The Query Handler parses the query and extract the where condition attribute and passes it to Computation Agent (CA)	\\
Step3:CA request the shares of the condition attribute from DB Server manager which forwards the request to all share holding databases\\
Step4:After getting the attribute shares, CA reconstruct the attribute values and check the condition and finds out the index values of satisfying attributes\\
Step5:Query Handler gives the select attribute name and request the CA to get the shares corresponding to the index values obtained in step 4\\
Step6:CA forwards a packet containing the following fields to
		\begin{table}[!h]			
				\caption{Packet Format}
				% title of Table
				\centering
				\begin{tabular}{|c|c|c| } \hline
				{\bf Indexvalue } & {\bf  Select attribute name} & {\bf  Client IP address} \\ \hline
				% used for centering table
		     	 \end{tabular}
			 	 \label{table:fields}
			 	 % is used
			 	 \end{table}
DB Manager and from there to database servers.\\	
Step7:The DB servers send the requested attribute column shares having 	the specified index values to the provided IP address of client\\
Step8:The result constructor in the client reconstructs the shares to retrieve the actual query result
\section{Example Scenario}	 	
Consider a hospital database system which contains patient's disease records. Since it is a large database, it is outsourced in a cloud storage. The database contains sensitive information so that the content of the database should not be revealed to a third party. And also suppose the hospital authority wants to know how many AIDS patients are there keeping the anonymity of the patient.
In this case, the hospital data is stored in 3 database servers in the form of shares and a (2,3) scheme is used. Each database item is formed in to shares using shamir's secret sharing scheme and stored in different servers having the same database schema.The database owner stores the data in the form of shares and Database Server Manager has the details of locations of database servers where these shares are getting stored.\\
%\section{Experimental Result}
Consider the patient\_details table as shown in Table \ref{table:patient database}.
	 		\begin{table}[ht]
	 		\small
	 		
	 		\caption{patient\_details Table}
	 		% title of Table
	 		\centering
	 		\begin{tabular}{|l|c|c|c| } \hline
	 		{\bf Patientid } & {\bf  Patientname} & {\bf  Doctorid} & {\bf  Diagonosis}\\ \hline
	 		 101 & Ann & 51 & Aids\\
	 		 102 & Bony& 21 & Cancer\\
	 		 103 & Cara & 51 & Fever\\
	 		 104 & Dona & 26& Aids\\[1ex]
	 	
	 		 % [1ex] adds vertical space
	 		 \hline
	 		 %inserts single line
	 		 \end{tabular}
	 		 \label{table:patient database}
	 		 % is used
	 		 \end{table}
Each attribute value is divided into 3 shares using shamir's secret sharing scheme and is getting stored in three database servers as in Table \ref{table:share in server I}, Table \ref{table:Share in Server II}, Table \ref{table:share in Server III} respectively.
		\begin{table}[ht]
						\small
						
						\caption{Database Server I}
						% title of Table
						\centering
						\begin{tabular}{|c|c|c|c|c| } \hline
						{\bf Index } & {\bf Patientid } & {\bf  Patientname} & {\bf  Doctorid} & {\bf  Diagonosis}\\ \hline
						 
				1& 189 & 1115 & 321 & 1111\\
				2& 168 & 479 & 743 & 931\\
				3 & 236 & 2314 & 209 & 832\\
				4& 247 & 641 & 659& 120\\[1ex]
			
				 % [1ex] adds vertical space
				 \hline
				 %inserts single line
				 \end{tabular}
				 \label{table:share in server I}
				 % is used
				 \end{table}
			 \begin{table}[!h]
			
			 \centering
			 % used for centering table
			\small
			\caption{Database Server II}
								% title of Table
			\centering
			\begin{tabular}{|c|c|c|c|c| } \hline
			{\bf Index } & {\bf Patientid } & {\bf  Patientname} & {\bf  Doctorid} & {\bf  Diagonosis}\\ \hline
			 % inserts single horizontal line
			1& 325 & 789 & 510 & 210\\
			2 & 320 & 197& 408 & 1245\\
			3 & 543 & 980 & 201 & 911\\
			4 & 468 & 319 & 877& 319\\[1ex]
		
			 % [1ex] adds vertical space
			 \hline
			 %inserts single line
			 \end{tabular}
			 \label{table:Share in Server II}
			 % is used
			 \end{table}
			 	 \begin{table}[!h]
			 		\small
			 							
			 		\caption{Database Server III}
			 							% title of Table
			 		\centering
			 	\begin{tabular}{|c|c|c|c|c| } \hline
			 	{\bf Index } & {\bf Patientid } & {\bf  Patientname} & {\bf  Doctorid} & {\bf  Diagonosis}\\ \hline]
			 	 % inserts table
			 	 %heading
			 	
			 	 % inserts single horizontal line
			 	 1 & 509 & 865 & 712 & 2354\\
			 	 2 & 558 & 107& 501 & 123\\
			 	 3 & 1024 & 954 & 646 & 986\\
			 	4 & 479 & 310 & 981& 912\\[1ex]
			 
			 	 % [1ex] adds vertical space
			 	 \hline
			 	 %inserts single line
			 	 \end{tabular}
			 	 \label{table:share in Server III}
			 	 % is used
			 	 \end{table}
\begin{enumerate}
\item Client generates the query Select Patientname from patient\_details where Diagonosis='Aids'.	
\item The query is passed to Query Handler (QH) module. QH extracts the where portion and take the attribute Diagonosis and send to Computation Agent (CA)
\item CA forwards the attribute name to Database manager
\item Database Manager request the 'Diagonosis' column values from any 2 database servers as per the threshold.
\item On getting the request each database server replies by sending the shares of the requested attribute column(Diagonosis) to Database Manager.
\item Database manager forwards the shares to CA where CA gets the Diagonosis column shares from any two of the database servers as shown in Table \ref{table:shares of diagonosis}
	\begin{table}[!h]
			\small
			
			\caption{Share from servers}
			% title of Table
			\centering
			\begin{tabular}{|c|c| } \hline
			{\bf Diagonosis } & {\bf  Diagonosis} \\ \hline
			% used for centering table
			
			% inserts table
			%heading
			\hline
		 	 1111 & 210 \\
		 	 931 & 1245 \\
		 	 832 & 911 \\
		 	 120 & 319 \\
		 \hline
		 	 \end{tabular}
		 	 \label{table:shares of diagonosis}
		 	 % is used
		 	 \end{table}
	 	
\item CA applies langrange interpolation to reconstruct the original values for Diagonosis field as shown in Table \ref{table:disease}
	\begin{table}[!h]
		 	 \caption{After Reconstruction}
		 	 % title of Table
		 	 \centering
		 	 % used for centering table
		 	 \begin{tabular}{|c|}
		 	 % centered columns (4 columns)
		 	  %inserts double horizontal lines
		 	\hline
		 	\bf Diagonosis\\ \hline 	 
		 	 Aids\\
		 	 Cancer\\
		 	 Fever\\
		 	 Aids\\
		 	 \hline
		 	 \end{tabular}
		 	 \label{table:disease}
		 	 % is used
		 	 \end{table}
\item CA checks the where condition of the query, that is diagonosis='Aids' and compute index values 1 and 4 which satisfies the condition
\item Query Analyser passes the select attribute name (patientname)	 to CA and CA forwards the packet containing index values, select attribute name, client IP address to Database Manager	
\item Database Manager requests the share holding DB servers to send the shares of attribute patientname in the specified indexes to the client IP address.
\item DB servers pass the shares to the specified IP address of the client
\item The result constructor in the client  receives all the three shares of patientnames as in Table  \ref{table:shares of patientname}
	\begin{table}[!h]
		 	 \caption{Shares of patientname}
		 	 % title of Table
		 	 \centering
		 	 \
		 	 % used for centering table
		 	 \begin{tabular}{|c|c|}
		 	 \hline
		 	\bf Patientname & \bf Patientname \\ \hline
		 	 	 
		 	 1115 & 789\\
		 	 641 & 319\\
		 	 \hline
		 	 \end{tabular}
		 	 \label{table:shares of patientname}
		 	 % is used
		 	 \end{table}
		\item After reconstruction, Client gets the result as in Table \ref{table:Result of Query}
		\begin{table}[!h]
			 	 \caption{Result of Query}
			 	 % title of Table
			 	 \centering
			 	 
			 	 % used for centering table
			 	 \begin{tabular}{|c|}
			 	 % centered columns (4 columns)
			 	  %inserts double horizontal lines
			 	\hline
			 	\bf Patientname\\[0.5ex]
			 	 % inserts table
			 	 %heading
			 	 \hline
			 	 Ann\\
			 	 Dona\\
			 	 \hline	 
			 	 \end{tabular}
			 	 \label{table:Result of Query}
			 	 % is used
			 	 \end{table}
\end{enumerate}			 	 
\section{Conclusion}	 
PIR and secure storage and retrieval of data in untrusted servers raise a major security challenge.
We presented a secure database storage and retrieval system based on secret sharing. Since the data is stored as shares in databases,  the knowledge of shares will not reveal any clue regarding the original data. The query analysis and result reconstruction are performed in the client side computation agent which ensures privacy preserving query processing and computation.The system proves to be efficient, secure and reliable. The work can be extended with unstructured data. The coalition of the service providers to retrieve the original data is a major security concern.A secret vector which contains the values used to evaluate the secret polynomial corresponds to each user can be used, which is known only to the clients and hence provides added security against untrusted service providers.Simple and efficient XOR based secret sharing scheme can be used, if the number of servers and  threshold is small.

\bibliographystyle{plain}
\bibliography{pir}

% that's all folks
\end{document}